\documentclass[aps,amsfonts,pra,twocolumn,showpacs]{revtex4-1}
\usepackage{epsfig,amsmath,amssymb,bm,epsf,graphics,psfrag,verbatim,subfigure}

\newcommand\beq{\begin{equation}}
\newcommand\eeq{\end{equation}}
\newcommand\bea{\begin{eqnarray}}
\newcommand\eea{\end{eqnarray}}

\begin{document}

\title{%
  Probing the mechanical properties of graphene using a corrugated elastic substrate
% Mechanical deformation of single- and few-layer graphene on micro-scale-grooved PDMS
}

\author{%
Scott Scharfenberg\rlap,$^{1}$
D.~Z.~Rocklin\rlap,$^{1,2}$
Cesar Chialvo\rlap,$^{1}$
Richard L.~Weaver\rlap,$^{1,2}$
Paul M.~Goldbart\rlap,$^{1,2}$
Nadya Mason\rlap,$^{1}$
}

\affiliation{%
$^{1}$Department of Physics and
$^{2}$Institute for Condensed Matter Theory,\\
University of Illinois at Urbana-Champaign,
1110 West Green Street, Urbana, Illinois~61801}

%\begin{abstract}
%We examine the mechanical properties of graphene samples of thickness ranging from 1 to 17 layers, placed on a %micro-scale-corrugated elastic substrate.  Using atomic force microscopy, we show that the graphene adheres to the %substrate surface, and can substantially deform the substrate, with larger graphene thicknesses creating greater %deformations.  We use linear elasticity theory to model the deformations of the composite graphene-substrate system.  %We compare experiment and theory, and thereby extract information on graphene's
%bending rigidity, adhesion energy, critical stress for interlayer sliding, and sample-dependent tension.
% [** NM: how about corrugated instead of grooved, throughout?]
%\end{abstract}

% \pacs{**}

\maketitle

%The exceptional mechanical properties of graphene have made it an attractive candidate for useful nano-mechanical and %electro-mechanical devices~\cite{Geimstatus}.  Two key aspects of its mechanical behavior are its elastic moduli and %adhesive capabilities.  The elastic properties have been measured using nano-indentation~\cite{Hone} and %pressurization~\cite{McEuen} techniques, and Young's modulus was found to be exceptionally high, $E\sim 1\,{\rm %TPa}$.  The van der Waals adhesion to surfaces has been examined theoretically~\cite{Sabio}, and local adhesion to %nanoparticles has been studied~\cite{Zong}.  However, it has typically proven difficult to extract experimental %parameters for adhesion, despite the fact that graphene's adhesive properties can strongly influence its electronic %and mechanical behavior. For example, substrate interactions highly modify graphene's doping~\cite{Andrei,Bolotin} %and carrier mobility.  Even suspended graphene is known to adhere to sidewalls, which introduces strain and modifies %mechanical behavior~\cite{Bunch,Metzger}. In addition, the mechanical interplay between graphene and other elastic %materials has not been well studied, although it is crucial to the use of graphene in composite~\cite{Geimstatus}, %flexible~\cite{Rogers}, or strain-engineered~\cite{Guinea} materials.
\textbf{
The exceptional mechanical properties of graphene have made it attractive for nano-mechanical devices and functional composite materials~\cite{Geimstatus}.  Two key aspects of graphene's mechanical behavior are its elastic and adhesive properties. These are generally determined in separate experiments, and it is moreover typically difficult to extract parameters for adhesion. In addition, the mechanical interplay between graphene and other elastic materials has not been well studied. Here, we demonstrate a technique for studying both the elastic and adhesive properties of few-layer graphene (FLG) by placing it on deformable, micro-corrugated substrates. By measuring deformations of the composite graphene-substrate structures, and developing a related linear elasticity theory, we are able to extract information about graphene's bending rigidity, adhesion, critical stress for interlayer sliding, and sample-dependent tension. The results are relevant to graphene-based mechanical and electronic devices,
% and to using
  and to the use of
graphene in composite~\cite{Geimstatus}, flexible~\cite{Rogers}, and strain-engineered~\cite{Guinea} materials.
}

The elastic properties of graphene have previously been measured using nano-indentation~\cite{Hone} and pressurization~\cite{McEuen} techniques, and Young's modulus $E$ was found to be exceptionally high, $\sim 1\,{\rm TPa}$.  Graphene's van der Waals adhesion to surfaces has been examined theoretically~\cite{Dunn}, and local adhesion to nanoparticles has been studied~\cite{Zong}.  Substrate interactions, due to surface adhesion, highly modify graphene's doping~\cite{Andrei,Bolotin} and carrier mobility.  In addition, adhesion to sidewalls in suspended
graphene introduces strain and modifies mechanical behavior~\cite{Bunch,Metzger}.
Here, we explore both elasticity and adhesion, which are evident in the interaction between micro-scale-corrugated elastic substrates and graphene samples of thickness ranging from 1 to 17 atomic layers. By using an atomic force microscope (AFM) to determine surface adhesion and deformations, we find that the FLG can fully adhere to the patterned substrate, and that thicker samples flatten the corrugated substrate more than thinner samples do. By developing a linear elasticity theory to model the flattening and adhesion as a function of layer thickness, we are able to extract estimates of, or bounds on, various fundamental and sample-dependent properties of the system.

%Here, we use an atomic force microscope (AFM) to probe the mechanical properties of graphene samples of %thickness ranging from 1 to 17 layers, placed on micro-scale-corrugated elastic substrates. We find that the FLG %can fully adhere to the patterned substrate, and that thicker samples flatten the corrugated substrate more than %thinner samples.  We develop a linear elasticity theory to describe the deformations of the graphene-substrate %composite structure, and combine experiments with theory to extract estimates of or bounds on multiple %fundamental and sample-dependent properties of the system.

\begin{figure}[hh]
\centerline{
\includegraphics[width=.45\textwidth]{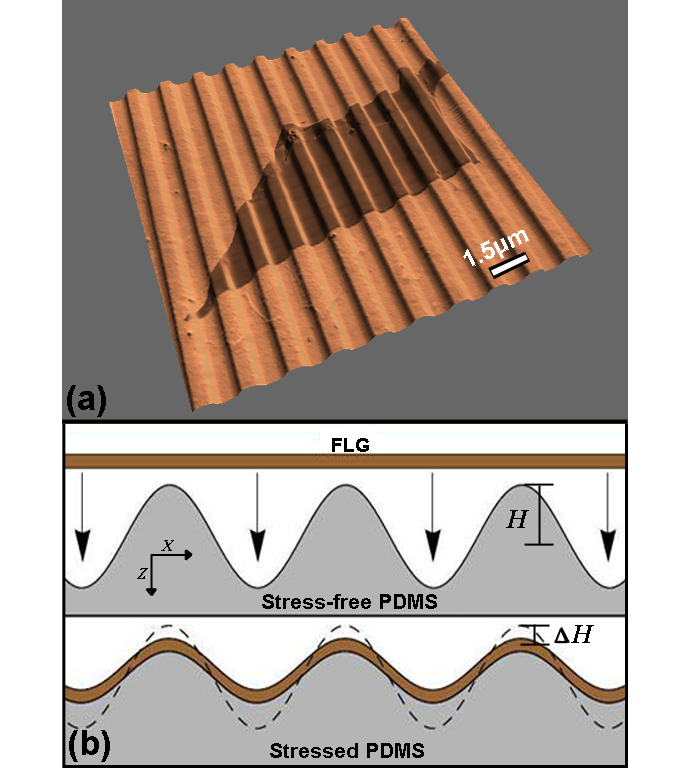}}
\caption{\label{Optical}
\small{{\bf a.}~AFM 3D topographic image of FLG on corrugated PDMS.
AFM height and phase data are superimposed to create color contrast.
% The color contrast is due to a superposition of AFM height and phase data.
{\bf b.}~Illustration of FLG-PDMS interaction, showing how FLG adheres to and flattens the PDMS corrugations.
The coordinates and amplitudes relevant to the theoretical model are labeled.}}
\end{figure}

%In this Letter, we report on experiments in which we deposit various thicknesses of few-layer graphene (FLG) on %to a micro-corrugated elastic material, and use an atomic force microscope (AFM) to probe the resultant %deformation of both the substrate and the FLG.  We find that the FLG can fully adhere to the patterned %substrate, and that thicker samples flatten the corrugated substrate more than thinner samples.  We develop a %linear elasticity theory to describe the deformations of the graphene-substrate composite structure.  Taken %together, our experiments and theory enable us to extract information on graphene's bending rigidity, adhesion %energy, critical stress for interlayer sliding, and sample-dependent tension.

\begin{figure}[hh]
\centerline{
\includegraphics[width=.48\textwidth]{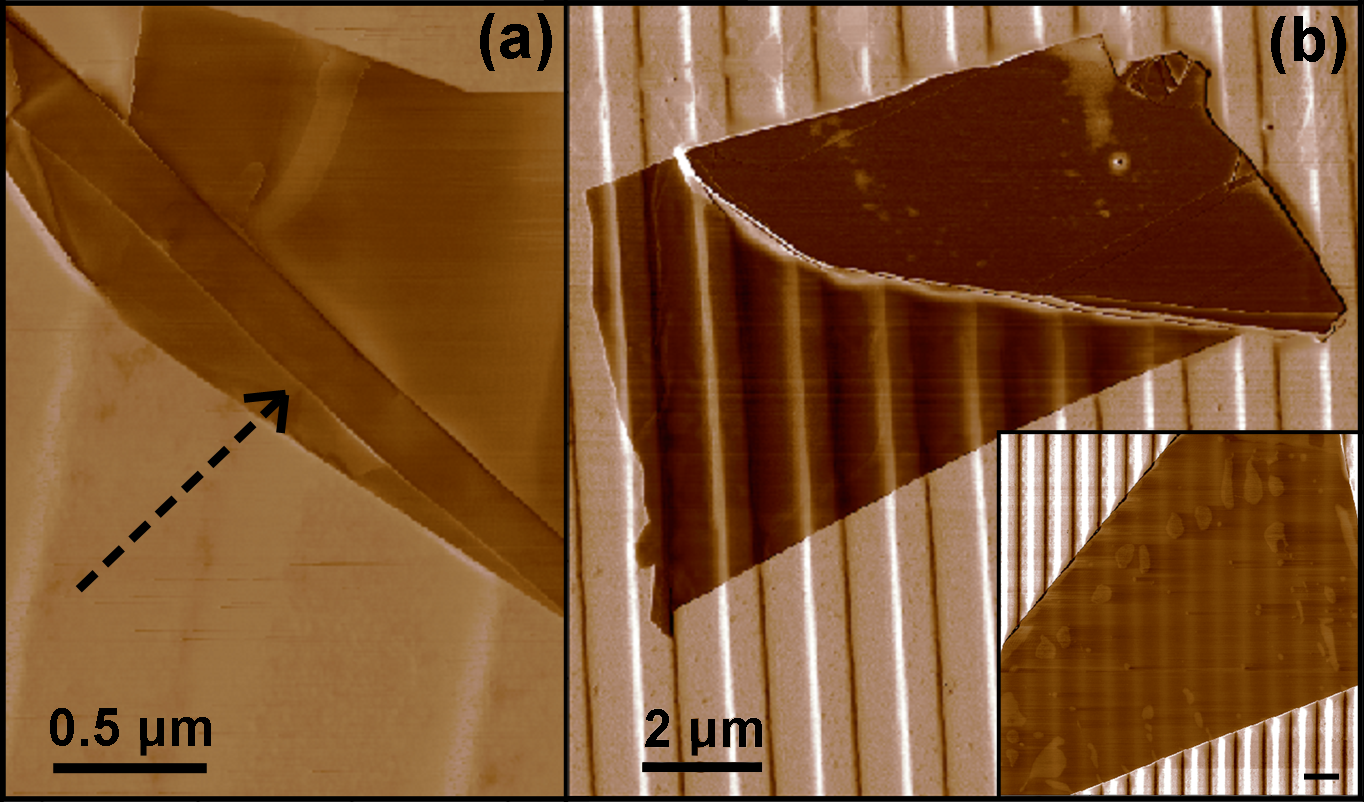}}
\caption{\label{phase}
\small{{\bf a.}~Micrograph of FLG folded by AFM tip.
FLG-FLG edges are used for thickness measurements.
Dashed arrow points to fold and shows the direction in which AFM tip was dragged.
{\bf b.}~AFM phase images of FLG on PDMS.
Main figure:~2-layer graphene (lower left section of FLG flake)
showing uniform adhesion over plateaus and valleys,
indicated by homogeneous contrast over them.
Adhesion loss, evident as bright and dark lines,
occurs only at steepest regions between plateaus and valleys.
Inset:~13-layer graphene showing inhomogeneous phase and adhesion;
bubble-like structure indicates where adhesion changes.
Scale bar is $2\,\mu{\rm m}$.
}}
\end{figure}

Sample substrates were prepared by casting a 3~mm thick layer of polydimethylsiloxane (PDMS)---which cures into a flexible, rubbery material---onto the exposed surface of a writable compact disc.
This resulted in approximately sinusoidal corrugations on the PDMS,
having a wavelength of $1.5\,\mu{\rm m}$ and a depth of $200\,{\rm nm}$ (see Fig.~1a).
Graphene was then deposited onto the PDMS via mechanical exfoliation~\cite{Scotch}.
Candidate samples were first located using optical microscopy, then imaged on an Asylum Research MFP-3D AFM.
Figure~1a shows a topographic image of FLG on the PDMS; it is evident from the image that the graphene conforms to the corrugations, as illustrated in Fig.~1b.

In order to fit the experimental data to a theoretical model, it was necessary to determine (1)~the thickness of the FLG, (2)~the adhesion between the FLG and the PDMS, and (3)~the height profile of the PDMS-FLG system.
The soft, non-standard substrate created difficulty in measuring FLG thickness via established AFM and Raman techniques. Thus, the thickness was determined
by using the AFM (in contact mode) to fold the flake onto itself, and then using the AFM again to measure the resultant FLG-FLG step height. An example of AFM-folded FLG is shown in Fig.~2a. The same flake could be folded at multiple locations to increase accuracy (typical accuracy was 1-2 layers), although, because the method was destructive, it had to be undertaken once all other measurements were completed.

The degree of adhesion between the FLG and PDMS was obtained by measuring the phase of the oscillation of the AFM cantilever.  This phase is determined by the electrostatic properties of the surface; in other words, sections having the same adhesion have common electrostatic properties and thus a common phase.  The main image of Fig.~2b shows a two-dimensional phase map for 2-layer FLG, in which the phase is uniform across the sample (except where adhesion is lost at the steepest slopes of the corrugation).
These data demonstrate the near-conformal adhesion between the FLG and the PDMS, and are consistent with previous work on graphene placed over more shallow depressions~\cite{Metzger}.  The AFM height data similarly indicate that the FLG adheres to the corrugations of the PDMS (e.g., see Fig.~3).  In contrast, the inset to Fig.~2b shows the phase data for 13-layer FLG, where bubble-like structure appears across the sample, showing that the phase is not uniform and, hence, that the FLG does not adhere well to the PDMS.  In general, we found that samples having fewer than ${\sim}11$ layers showed full adhesion, whereas thicker samples did not fully adhere. The adhesive properties did not seem to depend on the size or aspect ratio of the graphene samples, only their thickness.

\begin{figure}[hh]
\centerline{
\includegraphics[width=.5\textwidth]{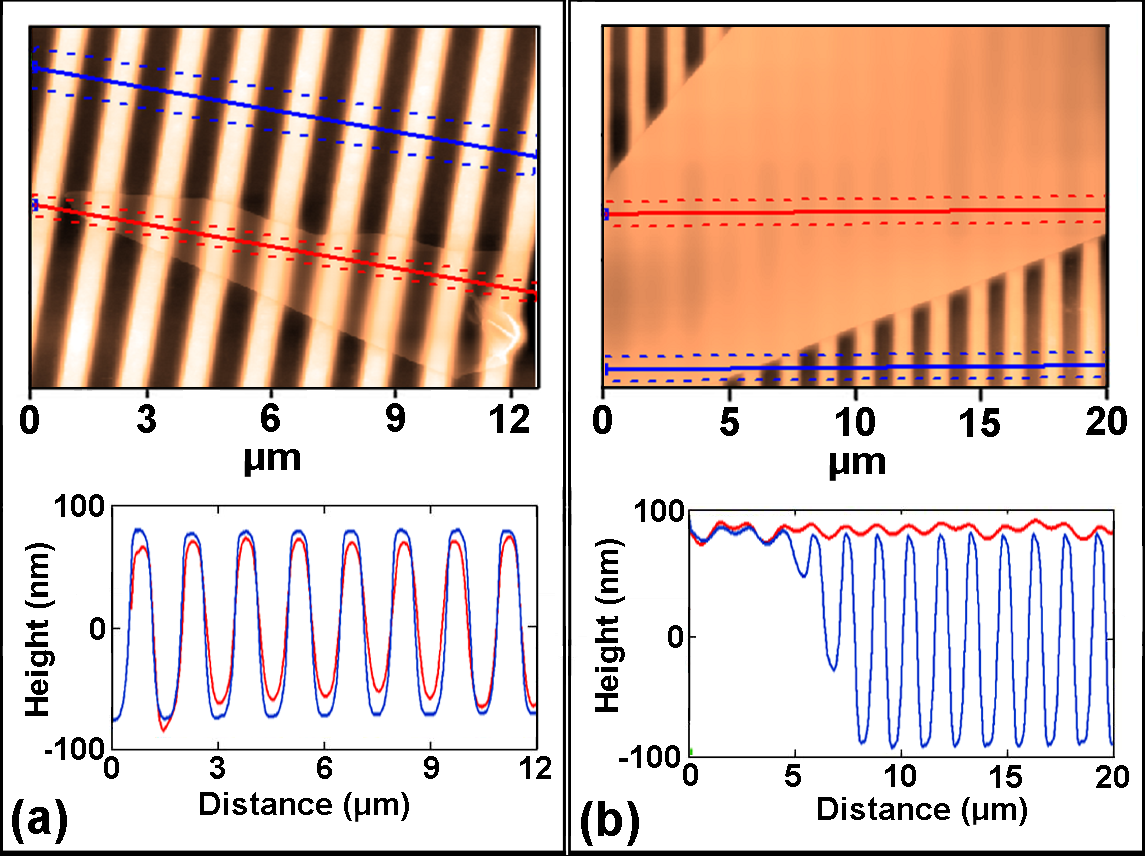}}
%\centerline{
%\includegraphics[width=.5\textwidth]{mgfigures/fig4.png}}
\caption{\label{Optical}
\small{Image (top) and height measurements (bottom) for
{\bf a.}~8-layer and
{\bf b.}~13-layer graphene.
Red lines show trajectories of scans over graphene, 
corresponding to red height curves 
(averaged between the dotted lines).
Blue lines show scans of surrounding PDMS substrate.
Scans over PDMS alone are taken far from FLG, to provide a baseline height unaffected by FLG.
Note the scale difference between vertical and horizontal axes in height vs.~distance curves.}}
\end{figure}

The most remarkable aspect of the FLG-PDMS system is that interplay between the rigidity of the graphene and the shear modulus of the PDMS causes the FLG to become corrugated and the PDMS to be flattened.  Figures~3a and b show image and height measurements for 8- and 13-layer FLG on PDMS, respectively.  In Fig.~3a it is clear that the FLG maintains the basic shape of the PDMS corrugations, but pulls the corrugations up in the valleys and pushes them down on the plateaus. In contrast, Fig.~3b shows that 13-layer FLG sits on top of the PDMS; while the FLG likely strongly deforms the PDMS, the amount of flattening is difficult to determine because of the lack of adhesion. Figure~4 shows the fractional height difference between the FLG-PDMS composite and the bare PDMS, plotted against graphene thickness, for 18 samples (measured on 9 different PDMS substrates); it is clear that the amount of flattening increases with layer number.

We now develop a linear elasticity theory that models the effect of stress at the interface between the relatively flat FLG
(ignoring the nanometer-scale intrinsic ripples)
%(We can safely assume that the intrinsic $\sim 1\,{\rm nm}$ amplitude ripples in graphene do not affect the FLG %response to
%the much larger PDMS corrugations)
%~\cite{NB-ripple-scale}
 and the corrugated PDMS.
% We shall then compare the results of the model to experimental data and hence determine
% graphene's bending stiffness as well as information about its interlayer cohesive and
% substrate adhesive properties.
%%%
To start, we consider an undeformed PDMS substrate having a corrugated surface: $h(x)=H \cos kx$ (see Fig.~1b for coordinates). Height profiles that are not simple sine waves can be handled via the superposition of suitable terms.  Graphene adheres to and flattens this surface, reducing the corrugation amplitude to $H-\Delta H$ (as discussed below).  The normal stress $S\cos kx$ required to deform the graphene in this way follows from thin plate theory~\cite{landau}, and is related to the deformed height profile via
\begin{equation}
M\nabla^{4}[(H-\Delta H)\cos kx]=S\cos kx,
\end{equation}
where $M$ is the flexural, or {\it bending\/}, rigidity of the FLG, so that 
$S=Mk^{4}(H-\Delta H)$.
An equal and opposite stress acts on the PDMS substrate, so we next determine how the PDMS responds to this stress.

% [** Can we match the figure fonts to the text fonts?]
\begin{figure}[hh]
\centerline{
\includegraphics[width=.5\textwidth]{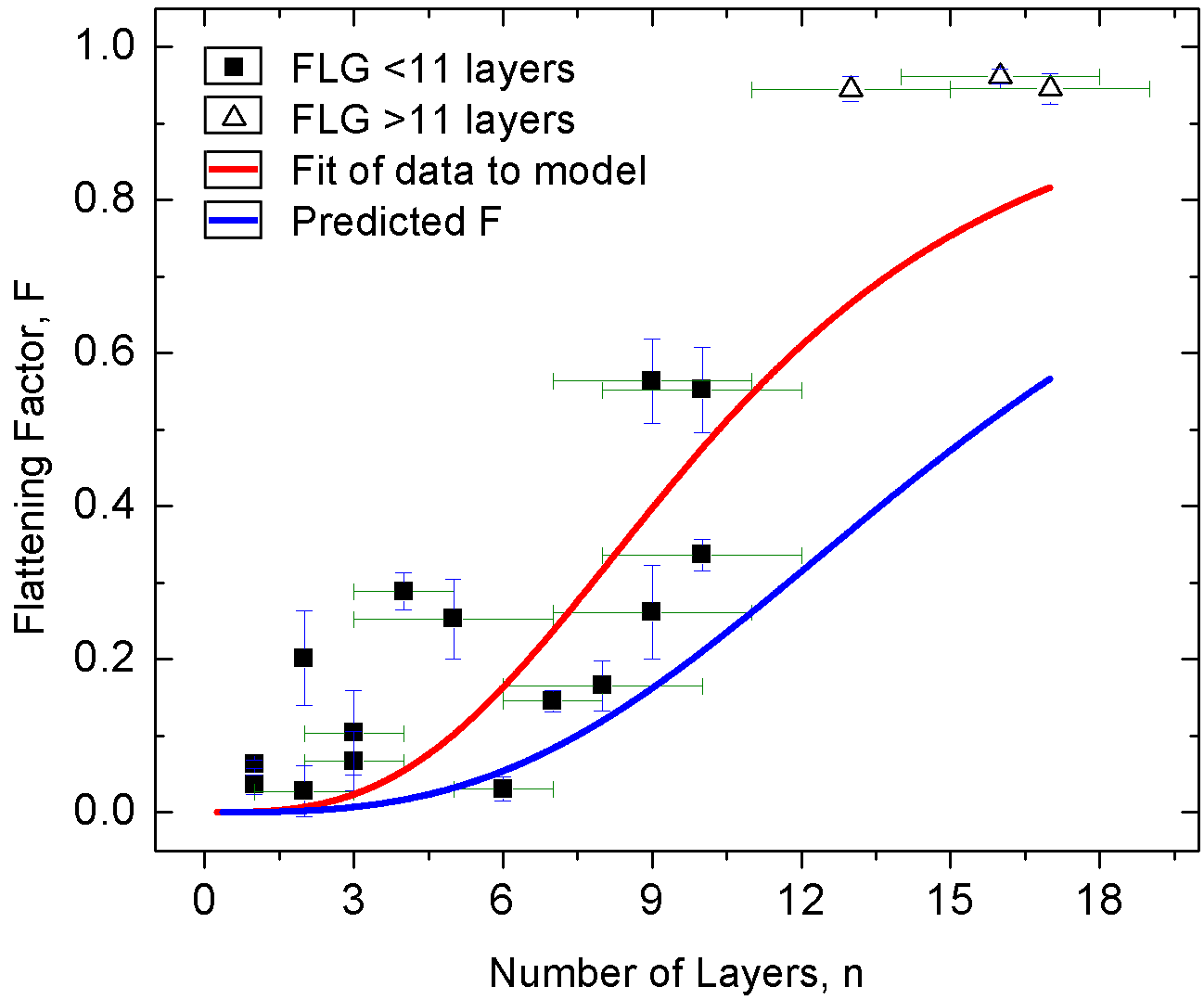}}
\caption{\label{Optical}
\small{
Data and fits for flattening factor (fractional deformation of PDMS) vs.~number of layers.  
Symbols show measured flattening for 18 FLG samples.
Samples with thicknesses $>11$ layers are shown with open triangles, as AFM height measurements are
likely modified by the lack of adhesion to the substrate.  
A value of $F= 0$ corresponds to no flattening
of the PDMS, whereas $F=1$ corresponds to totally flattened PDMS. Error bars are related to the uncertainty in layer number and the
spatial non-uniformity of flattening.  Red curve is least squares fit to model, assuming zero tension in samples.
Blue curve is predicted flattening for samples having zero tension (see text).}}
%
%
%Data and theoretical model fit for measured deformation $F$ vs.~number of Layers $n$ for 18 FLG samples.
%A deformation of $0$ corresponds to no flattening of the PDMS,
%whereas a deformation of $1$ corresponds to totally flat PDMS.
%Samples with thicknesses $>11$ layers are shown with open squares,
%as AFM height measurements were likely modified by the lack of adhesion to the substrate.
%Errors bars are related to uncertainty in layer number and non-uniformity of flattening.}}
\end{figure}

We regard the PDMS as being a semi-infinite, isotropic, incompressible medium, and describe distortions of it by means of a displacement field $\vec{u}(x,z)$.  
(We only consider configurations that are translationally invariant in the $y$-direction, and hence are effectively two-dimensional.)\thinspace\  
At the linear level, to which we are working, incompressibility implies divergence-free displacements:
$\vec{\nabla}\cdot\vec{u}=0$.
The displacement $\vec{u}$ obeys the condition of mechanical equilibrium, i.e., 
\begin{equation}
\label{Forces}
\mu\nabla^2\vec{u}=\vec{\nabla}p,
\end{equation}
in which $p$ is the pressure field, introduced to implement incompressibility, which requires that $\nabla^{2}p=0$.  In determining $\vec{u}$, the appropriate boundary condition involves specifying the normal component of the stress, which amounts to the condition (summing over repeated indices)
\begin{equation}
\mu\,
n_{i}\left(
\partial_i u_j\left(\vec{r}\right)+
\partial_j u_i\left(\vec{r}\right)
\right)-
n_{j}\,p\left(\vec{r}\right)=f_{j}\left(\vec{r}\right),
\end{equation}
in which $\vec{f}({ \vec{r}})$ is the external force acting on the PDMS at its surface and $\vec{n}(\vec{r})$ is the unit normal vector pointing outward from the PDMS surface.
The force exerted by the graphene on the PDMS can be taken to have a simple, sinusoidal form:
$(f_{x},f_{z})=(0,-S\cos kx)$.
It is then straightforward to show that the PDMS responds by undergoing the position-dependent displacement
\begin{eqnarray}
\left(\begin{matrix}
u_x \\ u_z
\end{matrix}\right)
&=&
-\frac{S}{2\mu k}{\rm e}^{-kz}
\left(\begin{matrix}
kz\,\sin kx\\ (1+kz)\cos kx
\end{matrix}\right).
% \\
% p&=&-S\,{\rm e}^{-kz}\cos kx.
\end{eqnarray}
In particular, the surface of the PDMS is displaced according to
$(u_{x},u_{z})\vert_{z=0}=-(S/2\mu k)(0,\cos kx)$,
which creates a surface profile $(H-\Delta H)\cos kx$. The amplitude is diminished from $H$ by an amount \begin{equation}
\Delta H=\frac{S}{2\mu k}=\frac{Mk^{4}(H-\Delta H)}{2\mu k}.
\end{equation}
Rearranging gives the {\it flattening factor\/}
\begin{equation}
F\equiv\frac{\Delta H}{H}=\frac{\phantom{1+}(Mk^{3}/2\mu)}{1+\left(Mk^{3}/2\mu\right)}.
\end{equation}%
A model of the FLG as a uniform elastic continuum suggests that $M=Dn^{3}$, where $n$ is the number of graphene layers and $D$ is the bending rigidity of a single graphene layer.  
We assume the thickness of $n$-layer graphene to be $n$ times the thickness of single-layer graphene.
Thus, we arrive at the result that $F$ should increase with increasing $n$.
%%%
%Next, we  determine how the graphene would distort in response to the opposing force.  As the graphene is thin, %compared with [**], we may apply the equilibrium condition for a plate subject to external forces~\cite{landau}, %i.e.,
%\begin{equation}
%D n^3\nabla^4 w(x)=-f_{z},
%\end{equation}
%where $w$ is the downward  displacement in response to the force $-f_{z}$,
%$D$ is the flexural rigidity of single-layer graphene,
%and $n$ is the number of layers.
%The appropriate displacement corresponding to the force $S\cos kx$ on the graphene  is
%$w(x,y)=(S/Dn^{3}k^{4})\cos kx$.
%For now, we regard the graphene sample as being a homogenous, isotropic elastic layer, in which case $D=Eh^3/{12%\left(1-\nu^2\right)}$, where $E$, $\nu$, and $h$ are, respectively, the graphene Young's modulus, Poisson ratio, %%and single-layer thickness.    For now, we also assume that any tangential components of the force that the graphene and PDMS %exert forces on one another are negligible.
%%%

We now compare the model to the data, to elucidate the FLG's mechanical behavior.
Figure~4 shows measured values of $F$ vs.~$n$,
along with a least squares fit to equation~(6) (red curve).
From the fit we extract a dimensionless graphene rigidity parameter
$G\equiv Dk^{3}/2\mu=0.00091$.
The shear modulus $\mu$ of PDMS was measured separately
via nanoindentation and an ultrasonic probe, which gave
$\mu=0.4\,{\rm MPa}$ and
    $0.23\,{\rm MPa}$,
respectively (the difference is possibly due to probing surface vs.~bulk moduli).
Using
$\mu=0.4\,{\rm MPa}$,
as well as
$k=4.2\,\mu{\rm m}^{-1}$,
the bending rigidity of graphene is then obtained as
$D=9.8{\times}10^{-18}\,{\rm Nm}$.
This value is higher than that predicted using Kirchhoff plate theory, from which 
$D=Et^{3}/12\left(1-\nu^2\right)=
2.9{\times}10^{-18}\,{\rm Nm}$,
using the graphene Young's modulus
$E\approx 900\,{\rm GPa}$~\cite{kelly},
Poisson ratio $\nu\approx 0$, and single-layer thickness
$t=0.335\,{\rm nm}$~\cite{kelly}.
The predicted values for $F$ are plotted as the blue curve in Fig.~3.

%\begin{eqnarray}
%G k^3=\frac{Eh^3 k^3}{24\mu(1-\nu^2)}
%\end{eqnarray}
%For $ k = 4.2$ $ \mu$$m^{-1}$, $h=.335\ nm$\cite{kelly}; $E\approx\ 500\ GPa$\cite{kelly}; $\nu=0$ a naive calculation of our %dimensionless parameter predicts $G k^3 \approx .00015$.

% The spread in the data exceeds that which can be accounted for by measurement uncertainty. This indicates that the different %mechanical behaviors in samples may originate from %the FLG deposition process.
%For thicker samples, there was a greater degree of uncertainty in the thickness measurements.  In regards to fitting %the model to the data, it was important to take this, as well as the uncertainty in the height measurements, into  %account when determining the most appropriate parameter values.

The spread in the data is greater than can be accounted for by measurement uncertainty.  
This leads us to hypothesize that the discrepancy between extracted and predicted values of $F$ (or $D$) is caused by tension in the graphene, resulting from sample-dependent friction between the PDMS and the graphene.  A tension $T$ could modify the flattening factor in equation~(6), giving
\begin{equation}
F\equiv\frac{\Delta H}{H_{0}}=
\frac{\phantom{1+}\left(Mk^{3}/2\mu\right) + \left(Tk/2\mu\right)}{1+\left(Mk^{3}/2\mu\right) + \left(Tk/2\mu\right)}.
\end{equation}%
If we assume that the difference between the predicted values of $F$ (blue curve) and the data in Fig.~4
is due to tension, we can use equation~(7) to extract a value of tension for each sample.
This yields tension values between $0$ and $0.20\,{\rm Nm}^{-1}$, with no discernible trend with thickness.
The tension is positive for each sample, to within the margin of error, consistent with the picture that friction opposes the contraction of FLG as it conforms to the corrugated PDMS.
% ~\cite{ref:ournote}

%%%
% which yields a range of T ~ xx for single-layer to T ~ xx for 9-layer graphene.
%%%
% This seems to be an orphaned snippet: which yields tensions per micron of FLG width of $1.47 \times 10^{-7}\,{\rm
% N}$.
%%%
% I don't like the 1 to 9 language since we don't seem to be sure what relation T has to n, and
% I don't like quoting a minimum since the lowest value is consistent with zero (and actually negative).
%%%
The maximum tension corresponds to a maximum axial strain
%%% 
%[**](fractional stretching of the graphene, where
%strain = $T/(E \times Area)$
%%% 
of $T/n h E=7.8\times 10^{-5}$.
We can also use the tension to estimate the magnitude of the stress due friction:
assuming the friction acts over a distance
$d \geq 10\,\mu{\rm m}$,
we find a  stress $T/d$ of less than
${2\times}10^{4}\,{\rm Pa}$. The condition that tension be positive, taken together with our data, implies that graphene's bending rigidity is no greater than $(1.6\pm0.8)\times 10^{-18}\,{\rm Nm}$, consistent with predicted values. If, as seems reasonable, the tension is negligibly small for at least one sample, then this bound would become an estimate of the graphene rigidity.

The data shown in Fig.~4 can also be used to estimate the normal interface stress $S=2\mu k F H$, which ranges  from  $(1.1 \mbox{ to } 3.0)\times 10^{5}\,{\rm Pa}$.
%%%
%Most values of the normal interface stress are substantially greater than the frictional stress, validating our %assumption that all forces on the system are normal forces.
%%%
The data also show that no samples which adhere to the surface have $F > 0.6$, implying that the adhesive strength between the graphene and the PDMS is $ \leq 3.0 \times 10^{5}\,{\rm Pa}$.
Note that this stress is model-independent.

We can extract bounds on the graphene-PDMS adhesion energy by considering
that the energy of the adhesion must be at least as large as the
spatially-averaged elastic deformation energy.
This energy can be regarded as a sum of contributions:
the elastic deformation of the substrate
$(1/2)\sigma_{i k} \big(u_{i k} -(p/\mu) \delta_{i k}\big) = S \Delta H \cos^2 (k x)/2$;
the FLG deformation under tension
$(T/2) (F h'(x))^2 =T(F H)^{2} k^2 \cos^2 (k x)/2$;
and the FLG bending
$(M/2)\big(F h''(x)\big)^2 = M(F H)^{2} k^4 \cos^2 (k x)/2$~\cite{landau}.
Ignoring the negligible tension contribution, these sum to $\mu k F H^{2}/2$.
The maximum elastic energy (which is also the lower bound of the adhesive energy) in our samples is thus
$0.044\,{\rm eV/nm}^2$.
This is consistent with the theoretical prediction of $0.04\,{\rm eV/nm}^2$ for the van der Waals adhesive energy between graphene and a $\mbox{SiO}_2$ substrate \cite{Sabio}.
% This is comparable to a typical value for a van der Waals energy.
Absent any other significant contributions to the energy budget (such as work done against friction), so adhesion energy must equal the elastic energy and $0.044\,{\rm eV/nm}^2$ becomes an estimate of the adhesive energy.

%%%
%As previously discussed, samples having thicknesses $> 11$ layers displayed ``bubble-like" phase structure and %nearly flat height profiles.
%If we conclude from this that the adhesion between the FLG and the substrate is on the order of the elastic energy %contained in rippled FLG of eleven layers, we have an adhesion energy of approximately $.04 \ J \ m^{-2}$.
%Taking the adhesion energy of thick samples to be the elastic energy of samples of $9\sim13$ layers we have adhesion %energy of $.3 \sim 10$ eV $/\buildrel _{\circ} \over {\mathrm{A}}^2$.  This is consistent with the prediction of %Sabio et al. of adhesion energy $.4$ eV $/\buildrel _{\circ} \over {\mathrm{A}}^2$ for graphene on a $\mbox{SiO}_2$ %substrate\cite{sabio}.
%%%

Because interlayer forces within FLG are weak, it is also important to consider shear forces, which could cause the graphene layers to slide relative to one another.  In this case, the impact on the flattening factor is to replace the bending rigidity of a cohesive sample $Dn^{3}$ by the sum $\sum\nolimits_{a}Dn_a^{3}$ of the bending rigidities $Dn_a^{3}$ of the individual slabs.  We find that such a model does not improve the fits to the data, and thus find no evidence that the graphene layers slide.  The physical effect of sliding would be to decrease the flattening factor, and thus sliding cannot explain the discrepancy between the theoretical values and data of Fig.~4.  We hypothesize that there does, however, exist some critical shear stress beyond which layers slide relative to one another. Considering the finite thickness of the FLG, Mindlin plate theory~\cite{Mindlin} shows that the boundary stress needed to deform the FLG generates a central shear strain $\epsilon$ of
\begin{equation}
\epsilon=n^{2}t^{2}k^{3}(H-\Delta H)/4.
\end{equation}
The absence of evidence for sliding in samples of $<11$ layers thus suggests a lower bound on the critical shear strain $\epsilon_{crit}$ of 
  $\geq 1.2 \times 10^{-5}$.
% $\geq 8.1 \times 10^{-6}$. [This value would go with a factor 1/6 in the epsilon equation.]
Multiplying by the graphene
shear modulus, which we take to be half its Young's modulus, 
this implies a critical shear stress between the layers of 
  $\geq 5.6\,{\rm MPa}$.

To conclude, we have developed a method of measuring the mechanical properties of graphene using deformable, micro-corrugated substrates.
% In the process, we have developed a technique for measuring the thickness of FLG flakes on soft substrates.
% For samples less than ten layers thick, FLG adheres to the substrate everywhere and deforms it.
% The flexural rigidity of the FLG competes with the stiffness of the substrate to produce flattened ridges.
% From the failure of adhesion at around 11 graphene layers, we have estimated a graphene-PDMS adhesion energy of %approximately $0.0070\,{\rm Jm}^{-2}$.   We have also found that the critical shear stress for interlayer sliding is %no less than $3.7\,{\rm MPa}$.
We are able to put bounds on---or extract estimates for---fundamental
properties such as graphene's bending rigidity, critical shear stress, and the FLG-PDMS adhesive strength and energy.  We also extracted sample-dependent properties such as the
tension and normal interface stress. 
The experimental and theoretical techniques developed in this paper may be readily extended to various substrates having a range of surface geometries. 
%%% 
% These techniques should lead to improved understanding and applications of mechanical 
% and electronic graphene-based devices, as well as graphene-based composite materials.
%%%

%flexural rigidity (bound, estimate if we assume at least one sample has negligible tension)
%adhesive energy bound, tentative estimate
%adhesive strength
%critical shear stress bound (definitely not estimate)

%axial strain of samples
%normal interface stress of samples
%tension of samples

\noindent\textit{Acknowledgments}.
We thank Scott Maclaren (UIUC CMM) and Richard Nay (Hysitron) for technical assistance.
This work was supported by
DOE DE-FG02-07ER46453,
NSF DMR06-44674,
NSF DMR09-06780,
and an NDSEG Fellowship (DZR).
%through the Frederick Seitz Materials Research Laboratory
%at the University of Illinois at Urbana-Champaign,

%\noindent\textit{Author Contributions.}
%S.S. prepared samples, carried out AFM measurements, and analyzed data;  
%C.C. assisted with sample preparation and design of experiment; 
%D.Z.R .and R.L.W. developed the theory and analyzed  data; 
%P.M.G. developed the theory; 
%N.M. conceived of and supervised the experiment.

\end{document}